\documentclass[%
preprint,
 amsmath,amssymb,
 aps,
pra,
floatfix
]{revtex4-1}
\usepackage{lineno}
\usepackage{graphicx}
\usepackage{dcolumn}
\usepackage{bm}
\usepackage{xcolor}
\usepackage{titlesec}
\titlespacing*{\section}{0pt}{0.1\baselineskip}{0.2\baselineskip}

\begin{document}

\title{Ultrafast dynamics of  three-dimensional Kane plasmons in the narrow-bandgap Hg$_{0.8}$Cd$_{0.2}$Te}

 \author{Xiaoyue Zhou$^1$}
 \author{Yi Chan$^1$}
 \author{Siyuan Zhu$^1$}
 \author{Fu Deng$^1$}
 \author{Wei Bai$^2$}
 \author{Jingdi Zhang$^{1,3}$}
 \email{jdzhang@ust.hk}
 \affiliation{
 $^1$Department of Physics, Hong Kong University of Science and Technology, Kowloon, Hong Kong SAR, China \\
 $^2$Key Laboratory of Polar Materials and Devices, Ministry of Education, Department of Electronics, East China Normal University, Shanghai 200241, China \\
 $^3$William Mong Institute of Nano Science and Technology, Hong Kong University of Science and Technology, Kowloon, Hong Kong, China
 }

\date{\today}

\begin{abstract}
We report on an ultrafast terahertz spectroscopic study on the dynamics of free carriers and the pertinent bulk plasmons in Hg$_{0.8}$Cd$_{0.2}$Te (MCT) film, a narrowband semiconductor accommodating three-dimensional massless Kane fermions. The ultra-broadband terahertz source enables the investigation of the lightly-doped equilibrium state in the presence of plasmon-phonon hybridization through the heavily-doped excited state, primarily dominated by plasmons. Without the recourse to the resource-consuming cryogenic high-magnetic field spectroscopy that hinges on observable related to the \textit{interband transition}, we show that the massless band dispersion can instead be conveniently perceived by the room-temperature study of the \textit{intraband transition} through the determination of the “plasmon–carrier density” relationship. We found the plasma frequency in MCT scales with the cube-root of carrier density ($\omega_p \propto n^{1/3}$), in contrast with the square-root scaling in the conventional massive fermion system of parabolic band dispersion. This work also answers the curious question of whether the MCT can maintain its massless Kane fermion character in case the strict gapless condition is deviated from. The method presented herein provides a convenient approach to identifying the landscape of both massless and massive band dispersion. 
\end{abstract}

\maketitle

\section{Introduction}

Electric charge carriers following a linear dispersion are normally known as relativistic fermions, owing to their negligibly small effective mass. This originated from the degenerate band touching points in the reciprocal space, including the Dirac point where inversion symmetry (IS) and time reversal symmetry (TRS) are preserved, and the Weyl point where either IS or TRS is broken. Neighboring these degenerate points, the quasi-particle excitations obey the massless Dirac equation \cite{bradlyn2016beyond,wehling2014dirac}, giving rise to the Dirac and Weyl fermions. Extensive investigations have been devoted to uncovering this linear dispersion of Dirac kind in various materials, including carbon nanotubes \cite{charlier2007electronic}, graphene \cite{RevModPhys.86.959}, topological insulators \cite{RevModPhys.82.3045}, 3D topological Dirac semimetals and Weyl semimetals \cite{PhysRevLett.108.140405,yang2014classification, PhysRevLett.107.127205, RevModPhys.90.015001}.

In systems of linear electronic dispersion, the massless quasiparticles dictated by the Dirac equation are mostly concomitant with zero energy gaps due to the abovementioned symmetry-imposed protections. There exists yet another school of materials with a strikingly small and tunable bandgap, in which linear dispersion prevails in the three-dimensional form but the associated massless quasi-particles disobey the Dirac equation; the massless carriers are therefore termed \textit{Kane fermions} for distinction \cite{orlita2014observation, teppe2016temperature, but2019suppressed}. Although the existence of the nontrivial Kane fermions with a 3D massless band dispersion was theoretically predicted decades ago \cite{kane1957band, zawadzki1974electron}, its experimental verification has only come about recently. The first identification of massless Kane fermions was made in the zinc-blende crystal Hg$_{1-x}$Cd$_{x}$Te (MCT) with a clear spectroscopic attribute of linear dispersion, despite the narrow but still-finite gap \cite{orlita2014observation}. It remains technically demanding to unequivocally verify 3D Kane fermions inhabiting a solid state system, on account that cryogenic temperatures and strong magnetic fields are necessary for spectroscopic determination of the band dispersion geometry. Unlike the massless fermions protected by symmetry, the linear dispersion in the Kane fermions in MCT is correlated to a shrinking energy gap, of which the size can be fine-tuned by varying the cadmium content to approach a critical level or by adjusting temperature \cite{hansen1982energy,teppe2016temperature}. An energy gap $E_g (x,T)$ in MCT amenable to external parameters therefore makes it a prominent semiconductor for optoelectric applications, such as infrared detectors \cite{rogalski2005hgcdte}. 
    
MCT undergoes a notable semiconductor-to-semimetal transition as the bandgap diminishes on reaching the critical value for either the dopant level or the base temperature. A bandgap highly sensitive to control parameters has made MCT the ideal platform for demonstrating quite some intriguing phenomena in the 3D Kane fermions, such as observation of linear dispersion of massless Kane fermions in MCT by broadband magneto-absorption spectra in the static state \cite{orlita2014observation},  2D collective Kane plasmon polariton in the heavily doped thin film MCT by ultrafast near-field microscopy in the mid-infrared spectral range (23-34 THz) \cite{charnukha2019ultrafast}. Nevertheless, the bulk plasmonic behavior of massless Kane fermions on the ultrafast timescale and at a milder doping level remain underexplored on account of (1) the limited detection range (0.1-3 THz) of conventional THz spectroscopy and (2) the intrinsic electron-phonon interaction in the 0.1-6 THz spectral range that complicates the extraction of the free-carrier and plasmonic responses. These limitations leave only the multi-THz spectral range (6-20 THz) the most suitable for uncompromised accuracy in parameter extraction. Technically, one, therefore, has to resort to novel THz sources for a simultaneous delivery of ultrafast temporal resolution and ultrabroadband detection range to evade the adverse effect from electron-phonon hybridization when determining the specifics of free carriers and the bulk plasmons, i.e., the plasma frequency $\omega_p$.

\section{Experiment}

The MCT film being measured in this work is 2 µm thick Hg$_{0.8}$Cd$_{0.2}$Te grown on semiconducting GaAs substrate using the molecular beam epitaxy (MBE) processes \cite{cao2025identification}. The choice of cadmium concentration x=0.2 is slightly above the critical concentration for a vanishing gap, so as to assure a THz conductivity minorly contributed by the thermally excited carriers and primarily by the photoexcited carriers in the non-equilibrium state.
 
To thoroughly explore the massless Kane fermion characteristics of MCT in the abovementioned terahertz frequency regime devoid of hybridization effect, we exploit the mid-infrared (MIR) pump-ultrabroadband THz probe spectroscopy to study ultrafast electrodynamics in the MCT flim on Hg$_{0.8}$Cd$_{0.2}$Te (x = 0.2), which is a narrow-bandgap semiconductor at or below room temperature (295 K) \cite{cook2000intense,hu2014optically,lan2019ultrafast}. The \textit{pump pulse} is generated from a tunable optical parametric amplifier (TOPAS-twins) driven by a Ti:sapphire laser amplifier (Coherent Astrella, center wavelength 800 nm, 35 fs), and the excitation wavelength is set at 2.4 µm for efficient carrier injection into the MCT film but none into the GaAs substrate. The \textit{probe pulse} is a near-single-cycle THz pulse spanning the 0.5-18 THz spectral range (2-72 meV), generated from a two-color laser-induced air-plasma filament. For exhaustive measurement of the complex dielectric function $\tilde{\varepsilon}(\omega)$, we perform the THz time-domain spectroscopy (THz-TDS) to map out, in full, the waveform of THz pulses by the electro-optical sampling technique with GaP (110-cut) and GaSe (z-cut) detection crystals to cover the 2-7.5 THz and 9-18 THz range, respectively. 

The THz pulses from the air-plasma source are reflected at 45-degree incidence either by the sample (MCT film on GaAs) or by the standard reference (gold). The reflected pulses are then measured either in the single-pulse scheme for sensing the equilibrium state or in the pump-probe scheme for probing the non-equilibrium state. We first Fourier transform the THz waveforms from the sample (MCT film) and reference (gold reference) scans, and then divide one with the other for $\tilde{r}(\omega)$, the complex reflection coefficient of the sample to output both magnitude and phase spectra. By varying the pump-probe delay $t_{pp}$, the dynamic reflection spectra $\tilde{r}$ ($\omega$,t$_{pp}$) can be measured at a sub-picoseond temporal resolution, which will be the primary input parameter for extraction time-dependent response functions $\tilde{\varepsilon}$ ($\omega$,t$_{pp}$) or $\tilde{\sigma}$ ($\omega$,t$_{pp}$) to be detailed in the following section and the Supplementary Information. Disimilar to classical materials that follow the parabolic single-particle dispersion $E(k)=\hbar^2 k^2/2m$ and conform to a dimension (\textit{D})-independent relation between the bulk plsamon frequency and the carrier density ($\omega_{p,D}\propto\sqrt{n_D}, D=1,2,3$), bulk plasmon frequencies in materials of massless single-particle dispersion—Dirac, Weyl and Kane fermions—are dimension sensitive and scale with the carrier density as $\omega_{p,D}\propto {n_D}^{D-1/2D}$ in the long-wavelength limit \cite{das2009collective}. The diametrically different scaling therefore suggests an experimental method to manifest the massless 3D Kane fermions by tuning the carrier density and tracking its scaling law with the bulk plasmon frequency; both are attainable by our ultrafast broadband THz spectroscopy.

\section{Results and Discussion}

Here, we present our findings on the ultrafast dynamic study of the narrow-bandgap Kane fermion semiconductor, MCT film on GaAs substrate. The broadband electrodynamic response of the sample in both equilibrium and non-equilibrium states is measured through complete sampling of the time-domain waveforms of the THz probe pulses at variable delay with respect to the excitation pump pulse. We obtain the raw complex reflection spectra from the above-described sample-reference scans and Fourier transformation of the time-domain waveforms and subsequently extract the dielectric function that bears much information on free carriers, infrared-active phonons, as well as their interplay in terms of the non-equilibrium dynamics upon pulsed photoexcitation. To explicitly investigate the free-carrier dynamics in the MCT film, the ideal excitation pulse is only capable of injecting carriers into the MCT layer but not into the underneath GaAs substrate. Therefore, the pump beam is set to center at 2.4 µm in wavelength (0.517 eV in photon energy), ensuring it stays above the bandgap of the MCT (0.155 eV) \cite{rogalski2005hgcdte} for efficient injection through linear absorption and well below that of the GaAs (1.43 eV) for fully suppressed absorption through either linear or two-photon processes. 

We start our investigation by performing broadband THz-TDS experiments on the equilibrium state of Hg$_{0.8}$Cd$_{0.2}$Te at room temperature. The time-domain waveforms reflected off the sample and the reference (gold) are Fourier transformed and divided for us to obtain the complex reflectivity spectra of the MCT film in the spectral range from 2.5 to 18 THz, as shown in Fig. 1(c). Note that there exists a spectral gap owing to the limited detection range of the EO-sampling crystals (GaP and GaSe) in our experiments, and such spectral gap coincides with the Reststrahlen band of unity reflection in GaAs (7-9 THz), resulting in little adverse effect on the extraction of key spectroscopic information. By employing the transfer matrix method \cite{born2013principles, yeh2006optical} built into a thin-film model of the unperturbed MCT layer on the thick semiconducting GaAs substrate, we extracted the complex dielectric function, $\tilde{\varepsilon}(\omega,0)$, of the static state Hg$_{0.8}$Cd$_{0.2}$Te, as shown by solid lines in Fig. 1(d). Further details regarding the analysis can be found in the Supporting Information. The static-state dielectric function at room temperature can satisfactorily fit to a multi-oscillator model that contains (1) a Drude component centered at zero frequency owing to the thermally excited carriers, $\omega_{p_0}$=2.81 THz, (2) the mode due to plasmon-phonon coupling at 3 THz, (3) HgTe-like transverse-optical (TO) phonon mode at approximately 3.7 THz and (4) the CdTe-like TO phonon at about 4.8 THz. In the high-frequency range (9-18 THz), we observed no other optical phonon modes. These observations on both free-carrier and phonon response of our Hg$_{0.8}$Cd$_{0.2}$Te film agree very well with previous studies \cite{PhysRevB.82.014306,polian1976dielectric,chu1993study,sheregii2009temperature}. Above the plasma frequency of 2.81 THz, we found that the dielectric function of the sample is primarily of dielectric and vibrational character in its equilibrium state. The multi-oscillator model for capturing vibrational and free-carrier response shall follow the Drude-Lorentz form below in order to reproduce the experimental results on complex dielectric function:
\begin{equation}\label{equation1}
    \varepsilon(\omega,0)=\varepsilon_\infty-\frac{\omega_{p_0}^2}{\omega^2+i\gamma_D\omega}+\sum_{k=1}^N \frac{S_k\omega_{TO,k}^2}{\omega_{TO,k}^2-\omega^2-i\gamma_{TO,k}\omega},
\end{equation}
where $\varepsilon_\infty$ denote the dielectric function due to all the high-frequency optical transitions out of our detection range, $\omega_{p_0}$ plasma frequency of the thermally excited free cariers, $\gamma_D$ free-carrier scattering rate, $S_k$ spectral weight (strength) of the k-th Lorentz oscillator, $\omega_{TO,k}$ center frequency of the $k_{\mathrm{th}}$ oscillator and $\gamma_{TO,k}$  damping rate, respectively. The fit of the phonon spectra of Hg$_{0.8}$Cd$_{0.2}$Te in the static state to the model (dashed lines) are shown in Fig. 1(d) and satisfactorily agrees with the experiment (solid lines). A total of four oscillators (1 Drude, 3 Lorentz) are employed, and their details are provided in Table 1.


To investigate the ultrafast carrier dynamics, the pulsed MIR excitation pulse (center wavelength 2.4 µm, pulse duration 100 fs) is used to initiate interband transition in the MCT film, and the broadband THz probe pulse at a variable delay provides insights into the time-dependent (dynamic) response function of the sample from 2.5-18 THz (see Fig. 1(a) for a schematic of the experiment). As IR-active phonons only exist at frequencies of 5 THz and below, the broad detection range allows one to investigate the bulk plasmons in terms of reflection spectra not only in the regime of relatively low carrier density that promotes strong plasmon-phonon coupling, but also offers access to the regime of high carrier density with negligible effect from optical phonons. The electronic band structure near the $\Gamma$ point of this 3D system governs whether one should expect the classical fermions promoted by trivial parabolic dispersion or the novel Kane fermions prompted by a conical (massless) dispersion, as a result of spin-orbit coupling, as illustrated in Fig. 1(b). Within the Kane fermion regime, the light-induced interband transition should create electrons in the conduction band and, in general, holes in the degenerate light-hole (LH) and heavy-hole (HH) bands, respectively. However, the HH-to-conduction-band pathway has the dominance in photo-doping process, resulting from the higher joint density of states for the transition related to the HH band than to the LH band \cite{orlita2014observation}. Furthermore, on account of the enormous difference in effective mass between that of electrons in the conical conduction band and that of heavy holes in the HH band, the plasma frequency $\omega_p$ shown in our THz spectra can exclusively be attributed to massless Kane electrons in the conduction band, and irrelevant to the heavy holes. We, therefore, take the spectroscopic results on the bulk plasma frequency $\omega_p$ as the primary indicator of the Kane fermion dynamics. Its value can be qualitatively identified by the reflection edge in the raw spectral or quantitatively determined by a two-step parameter extraction; first with the transfer-matrix method for dielectric function, and then the Drude-Lorentz multi-oscillator model.

The resultant ultrafast dynamics of the spectroscopic reflection coefficient is shown in Fig. 2(a) at various pump-probe delays $t_{pp}$ (vertical axis) with MIR excitation at the fluence of 16 $\mu$J/cm2, an intensity far below that for a pump-probe signal saturation, normally as the result of the bleached joint density of states for the inter-band transition. We observe in the time-resolved reflection coefficient that it takes approximately 2 ps for the photo-induced carrier proliferation to complete and reach a maximum value, and about 40 ps for the excited carriers to relax and recombine. Notably, the surging carrier density (equivalently the plasma frequency) gives rise to the predominance over the reflection coefficient by imposing a global increase toward the unity (|r|=1) across nearly the entire detection range of 2.5-18 THz, and it brings about a strong screening effect on the phonon modes, leading to the less pronounced vibrational feature. Further insights into the non-equilibrium state of MCT may be attained by the transfer matrix method (Supporting Information) that extracts the spectroscopic complex dielectric function, of which real and imaginary parts ($\varepsilon_1$, $\varepsilon_2$) are respectively shown in Fig. 2(c) and (d). On the arrival of the pump pulse, the real dielectric function $\varepsilon_1$ suddenly reverses its sign from positive to negative and substantially increases in magnitude, of which the impact extends up to around 20 THz. Concomitantly, the imaginary part $\varepsilon_2$ in the excited state is joined with an overwhelmingly strong component that scales inversely with frequency, in stark contrast to that in the equilibrium state in the presence of strong phonon response but much weaker bulk plasmonic response. The striking change in real and imaginary dielectric function jointly signifies such dominance of an excessive amount of free carriers as is pronounced by the THz frequency optical response. 

As the photoexcited holes primarily populate the heavy hole (HH) band and have an effective mass approximately 80 times greater than light electrons in the conduction band, the detected free-carrier response is governed mainly by massless Kane fermions in the conduction band and negligibly by light and heavy holes. These photo-excited carriers result in a dynamic contribution to the complex dielectric function and can be well captured by incorporating an additional Drude component into the oscillator model in Eq. 1\cite{dressel2002electrodynamics}, 

\begin{equation}\label{equation2}
    \varepsilon(\omega,t_{pp})=\varepsilon(\omega,0)-\frac{\omega_{p_{ex}}^2(t_{pp})}{\omega^2+i\gamma_{D_{ex}}\omega}
\end{equation}
where $\omega_{p_{ex}}$ and $\gamma_{D_{ex}}$ denote the plasma frequency and the scattering rate related to the photoexcited carriers. At the fluence of 16 $\mu$J/cm$^2$, the dynamic reflection coefficient can be best fit by setting $\omega_{p_{ex}}$ to exponentially decay as a function of pump-probe delay (lifetime 25 ps) from a maximum 19 THz and $\gamma_{D_{ex}}$ to be at 3 THz, as shown in Fig. 2(e).

To distinguish the electronic band dispersion of Kane fermion characteristic from that of a conventional narrow-bandgap semiconductor, we resort to the excitation fluence (\textit{F}) dependence of the peak amplitude in dynamic plasma frequency ($\omega_{p_{ex}}$). On account that, in the perturbative regime, the excitation fluence \textit{F} scales linearly with the photoexcited carrier density $n$, the power law for plasma frequency on fluence ($\omega_{p_{ex}}\propto F^\beta$) equivalently registers that on the carrier density $\omega_{p_{ex}}\propto n^\beta$. At a fixed excitation photon energy, a higher excitation fluence will populate the conduction band with increasingly more electrons. The exact form of the power law, i.e., value of the exponent, carries decisive information on the electronic dimensionality; particularly, on whether the pertinent band dispersion is classical (near parabolic) or massless (Dirac, Weyl, and Kane type) \cite{das2009collective}.  In practice, details about the time- and fluence-dependent plasma frequency $\omega_{p_{ex}}$ can be qualitatively identified from the plasma edge appearing in the ultrabroadband THz reflection spectra and quantitatively determined by a fit to the abovementioned multi-oscillator model that accounts for, in full, contributions by free carriers and optical phonons. Considering that the much lower carrier concentration ($3.5\times10^{16}$cm$^{-3}$, 295 K) in the static state \cite{madarasz1985intrinsic} than in the excited state by orders of magnitude, the non-equilibrium-state plasma frequency $\omega_{p_{ex}}$ can be primarily attributed to the photoexcited carrier. Fig. 3(a) shows the experimental results on broadband spectra of THz reflection coefficient at various excitation fluence (0-16 $\mu$J/cm$^2$) at a fixed pump-probe delay of 2 ps, at which the carrier density reaches its maximum and dynamic plasma frequency $\omega_{p_{ex}}$ can be determined at different fluences. The effect of the varying carrier density, or dynamic plasma frequency $\omega_{p_{ex}}$, on the reflection coefficient can directly be reproduced by numerical simulation (Fig. 3(b)) through embedding the dielectric function (Eq.\ref{equation1} and Eq.\ref{equation2}) into the transfer matrix model. It is with the same fluence-dependent plasma frequency, employed in numerical calculation of reflection coefficient, that we plot in Fig. 3(c) and observe the plasma frequency scale as the cube-root of the carrier density ($\omega_{p_{ex}}\propto F^{1/3}\propto n^{1/3}$) consistent with the 3D massless fermion, as opposed to a square-root dependence ($\omega_{p_{ex}}\propto F^{1/2}\propto n^{1/2}$) in the classical semiconductor of a parabolic band dispersion \cite{das2009collective}. To benchmark our experiment and conclusion on the 3D Kane fermion system, we performed the same experiment and analysis on the classical semiconductor film InAs on GaAs in a nearly identical configuration. The scaling law measured from InAs film is found to follow the square-root dependence ($\omega_{p_{ex}}\propto n^{1/2}$) (see Supporting Information) and, therefore, reaffirms the accuracy of the methodology described herein.

\section{Conclusion}

In conclusion, this work focused on the study of ultrafast dynamics of the bulk plasmons in the narrow-bandgap semiconductor Hg$_{0.8}$Cd$_{0.2}$Te that features the 3D Kane electronic band dispersion. We showed that, at ambient conditions and using the ultrabroadband THz pump-probe spectroscopy, one could precisely determine the power law for the plasma frequency and the carrier density. The insight into the shape of the band dispersion is made possible by the novel ultrabroadband THz source of pump-probe capability and facilitated by quantitative parameter extraction with the Drude-Lorentz model augmented by the transfer matrix method. Reliability of the approach is examined by comparison in the power law between the MCT/GaAs and InAs/GaAs samples; the former shows a cube-root dependence ($\omega_{p_{ex}}\propto n^{1/3}$) and the latter a square-root dependence ($\omega_{p_{ex}}\propto n^{1/2}$). The method reported herein is key to the identification of the 3D Kane fermion in the context of Hg$_{1-x}$Cd$_{x}$Te and is applicable to other massless fermionic systems in a more general context. 

\section*{Funding} This work was supported by National Key Research and Development Program of China (Grants No. 2020YFA0309603); National Natural Science Foundation of China (No. 12122416); Hong Kong Research Grants Council (Project No. ECS26302219, GRF16303721, GRF16307124, GRF16306522).



\begin{figure}[htbp]
\centering\includegraphics[width=0.9\textwidth]{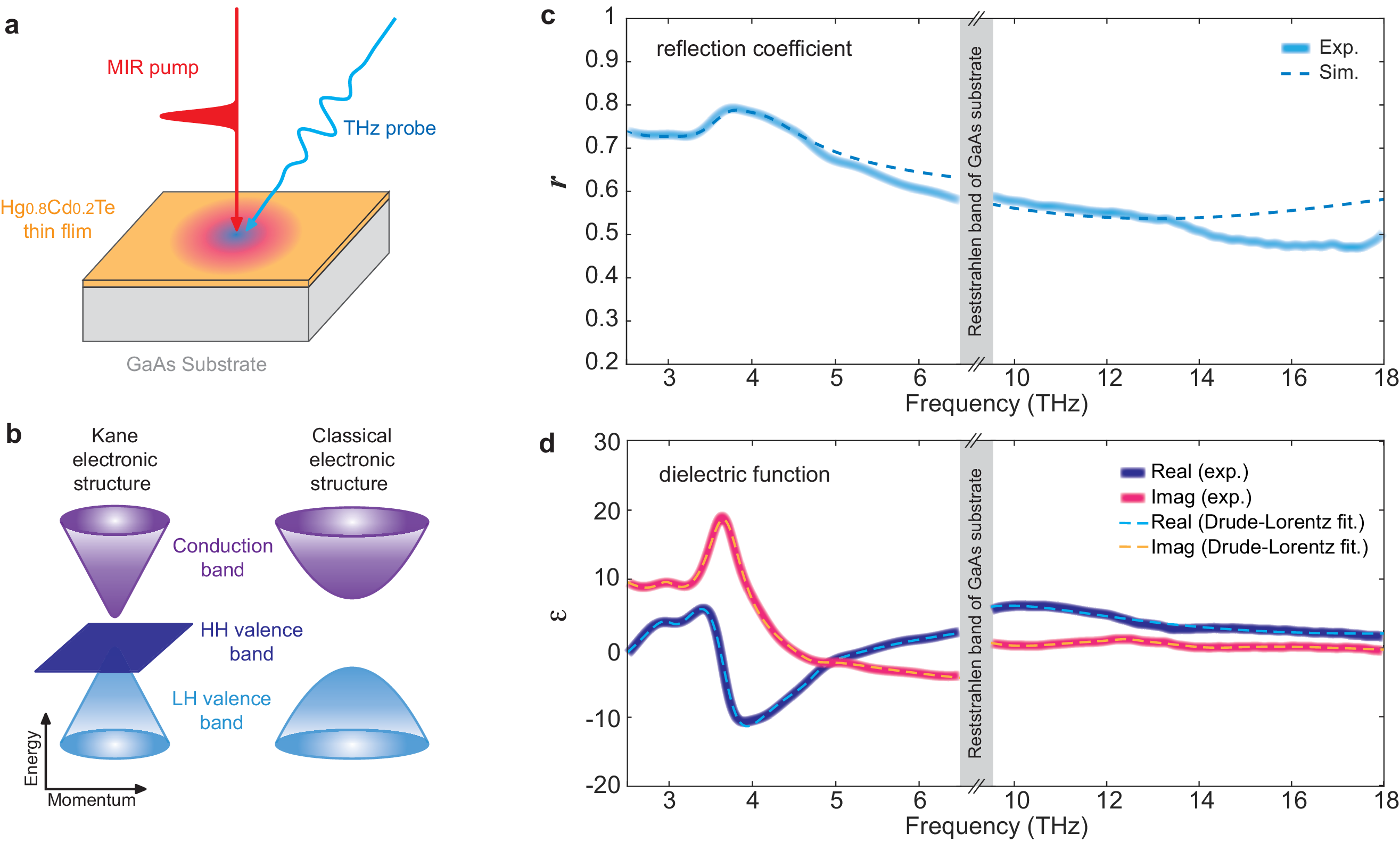}
\caption{Equilibrium optical response functions in MCT film and photo-doping experimental setup. (a) Schematic diagram of the MIR pump-THz probe experiment. (b) 3D Kane conical electronic band structure and classical parabolic electronic band structure near the $\Gamma$ point in the narrow-bandgap semiconductor. (c) Results from experiment (solid lines) and numerical simulation (dashed lines) on the amplitude of reflectivity, $|\tilde{r}|$, of the MCT film on GaAs substrate in the equilibrium state at room temperature. (d) Complex dielectric function, $\tilde{\varepsilon}$, of the MCT film extracted by the thin-film model (solid lines) and fit to a multi-oscillator model (dashed lines).}
\end{figure}

\begin{table}[ht!]
\caption{Parameters for the Drude-Lorentz oscillator}
  \label{tab:oscillator}
  \centering
\begin{tabular}{lrrr}
\hline
\hline
Drude & $\omega_{p_0}$ & $\omega_{0}$ & $\gamma_D$ \\
\hline
1 & 2.81 & 0 & 0.41 \\

\hline
\hline
Lorentz & $S_k$ & $\omega_{TO,k}$ & $\gamma_{TO,k}$ \\
\hline
$k = 1$ & 4.323 & 3.00 & 0.41 \\
$k = 2$ & 45.362 & 3.67 & 0.63 \\
$k = 3$ & 3.607 & 4.75 & 0.49 \\

\hline
\footnotesize
$^a$ all values are in the unit of THz.

\end{tabular}
\end{table}

\begin{figure}[ht!]
\centering\includegraphics[width=0.9\textwidth]{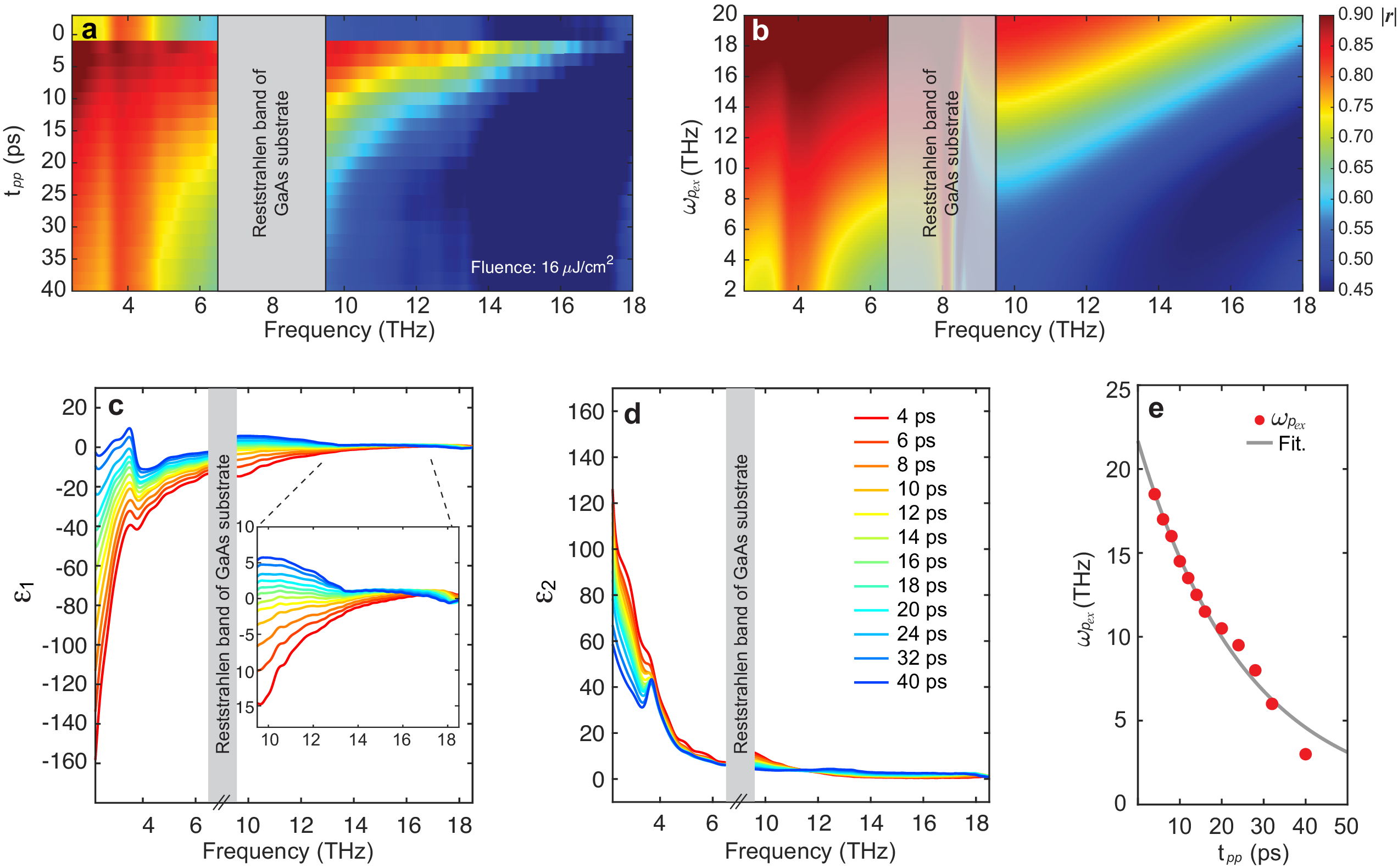}
\caption{Ultrafast photo-induced optical response in Hg$_{0.8}$Cd$_{0.2}$Te. (a) Experimental results on the time-dependent reflection coefficient. (b) Numerical simulation on the dependence of reflection coefficient on plasma frequency ($\omega_{p_{ex}}$). (c) Real part of the dielectric function $\varepsilon_{1}$ at various pump-probe delays. The inset shows a zoomed-in view of the spectral range between 9 and 18 THz, and (d) the imaginary part $\varepsilon_{2}$; all extracted with the transfer matrix method. (e) Exponential decay dynamics of the plasma frequency $\omega_{p_{ex}}$ of the photoexcited Hg$_{0.8}$Cd$_{0.2}$Te thin film.}
\end{figure}

\begin{figure}[ht!]
\centering\includegraphics[width=0.9\textwidth]{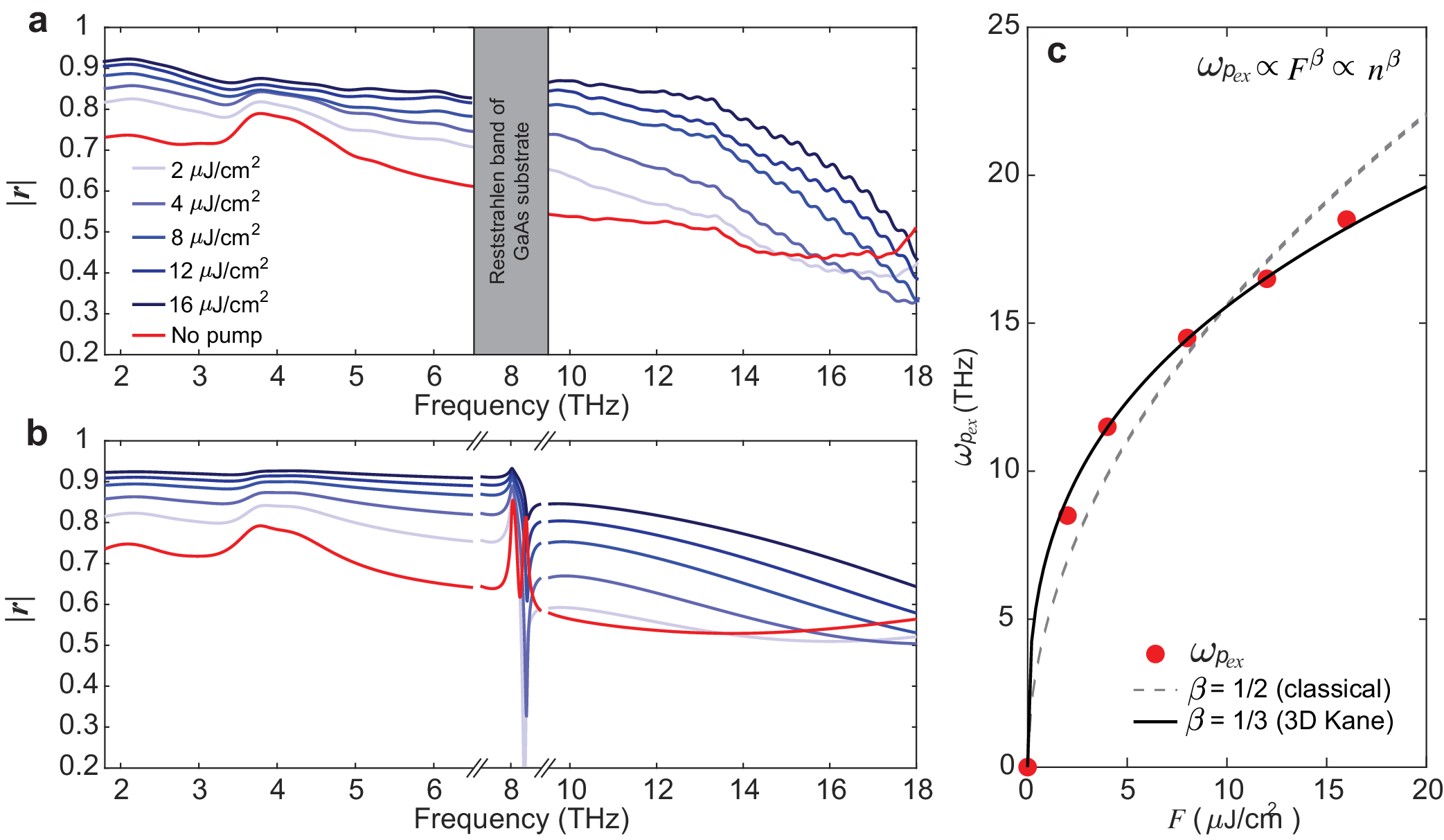}
\caption{Pump fluence dynamics at $t_{pp}$ = 2 ps. (a) Experimental and (b) simulated reflection coefficient, $|\tilde{r}|$, of Hg$_{0.8}$Cd$_{0.2}$Te thin film on GaAs substrate as a function of frequency and pump fluence. (c) Narrow-bandgap semiconductors of Hg$_{0.8}$Cd$_{0.2}$Te indicate the best fit (black solid lines) to the plasma frequency, $\omega_{p_{ex}}$ (red dots), obtained using the power-law with an exponent $\beta$ on pump fluence \textit{F} of either 3D Kane ($\beta$ = 1/3), or classical ($\beta$ = 1/2) electronic band structure.}
\end{figure}

\end{document}